\input{aipcheck}
\documentclass[
    ,final            
  ]
  {aipproc}

\layoutstyle{6x9}

\begin{document}

\title{The rate of type Ia Supernovae and the Star Formation History}


\author{L. Greggio}{
  address={INAF, Osservatorio Astronomico di Padova, Vicolo dell'Osservatorio 5,35127 Padova, Italy}
}

\author{E. Cappellaro}{
  address={INAF, Osservatorio Astronomico di Padova, Vicolo dell'Osservatorio 5,35127 Padova, Italy}
}
\begin{abstract}
The scaling of the rate of type Ia Supernovae (SNIa) with the parent galaxies'
color provides information on the distribution 
of the delay times (DTD) of the SNIa progenitors. We show that this information
appears to depend on the photometric bands used to trace the stellar age
distribution and mass-to-light ratio in the parent galaxies. Using both 
$U-V$ and $B-K$ colors to constrain
the star formation history, we model the SNIa rate as a function 
of morphological galaxy type for different DTDs. The comparison with the 
observed rate per unit $B$ and $K$ band luminosity yields consistent results, 
although
the large error bars 
allow us to exclude only very flat and very narrow DTDs. The number of
SNIa events per unit mass from one stellar generation results of 
$\sim$ 0.002-0.003 M$_\odot^{-1}$. 
\end{abstract}

\maketitle

\section{Introduction}

While it is generally believed that the progenitors of type Ia Supernovae 
(SNIa) are close binary systems,
the evolutionary path leading to the final explosion is still unclear. As 
a consequence, a variety of distributions of the delay time (i.e. the time 
between the birth of the binary and the SNIa event) is possible, according to
different scenarios for SNIa progenitors.
The distribution of the delay times (DTD) determines the rate of 
injection of energy and nucleosynthetic products from SNIa to the interstellar 
medium (ISM);
therefore, it is a key ingredient for modelling the evolution of galaxies,
as well as of the intracluster and intergalactic medium.
The interplay of the DTD and Star Formation History (SFH) for determining 
the SNIa rates in stellar systems has been studied by several authors
(\cite{Pilar},\cite{MDP},\cite{GR}). For a stellar system born  
according to a SF rate (SFR) given by $\psi(t)$, the SNIa rate at epoch $t$ is:
\begin{equation}
\dot{n}_{\rm Ia}(t) \, = \, k_{\rm Ia} \int_{t=0}^{t} {\rm DTD}(t_{\rm d}) \psi(t-t_{\rm d}) \,\,\, {\rm d}\, t_{\rm d} \, = \, k_{\rm Ia} \times M_{\rm SF} \times \langle {\rm DTD} \rangle_{\psi} 
\end{equation}   
where $t_{\rm d}$ is the delay time, $k_{\rm Ia}$ is the number of SNIa per 
unit 
mass in stars from one stellar generation, and $M_{\rm SF}$ is the total mass
transformed into stars in the system up to the current epoch $t$.
Therefore the time evolution of the SNIa rate, and/or the current rate 
in systems with different SFHs, can be used to constrain the shape 
of the DTD. 
At present, the cosmic evolution does not seem promising
in this respect, due to i) the smearing effect of the cosmic SFH, 
which results into a wide age distribution in systems at intermediate and low 
redshift; ii) the large error bars affecting the determination of the rate at 
the highest redshifts (\cite{Forster}, \cite{Blanc}, 
\cite{Botticella}). 
An alternative venue comes from the 
analysis of the SNIa rate in different galaxies:  
the  dependence of the rate from the age distribution of the
parent stellar population is sensitive to the shape of the DTD; in turn, the
age distribution can be traced using the colors of the stellar populations, so
that the quantitative variation of the specific SNIa rate with the color
of the parent galaxies constrains the DTD.
This is investigated in this paper, through modelling and comparison with the 
observations.
   
\section{DTDs and Star Formation Histories}

\begin{figure}
\includegraphics[height=.35\textheight]{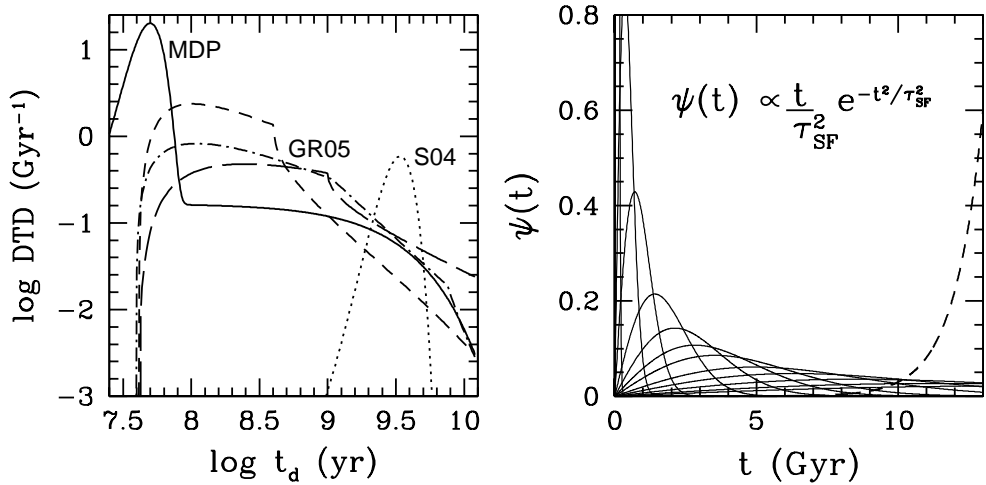}
  \caption{DTDs and SFRs used in this paper. Left: the MDP curve is 
constructed following the prescriptions in \cite{MDP}; the S04 
curve is a gaussian 
centered on $t_{\rm d}=$3.4 Gyr and with a width $\sigma=$0.68 Gyr. 
The GR05 curves include a steep
DD CLOSE (short dashed, $\beta_{\rm g}=-0.975, \tau_{\rm n,x}=0.4$ Gyr), an 
SD model with
a flat distribution of the mass ratios and a Salpeter distribution of the
primaries' masses (dot-dashed), and a flat DD WIDE (long dashed,
$\beta_{\rm a}=0, \tau_{\rm n,x}=1$ Gyr) models. See GR05 for the definition of
the parameters. Right: SFRs generated
by the labelled expression with $\tau_{\rm SF}$ varying between 0.1 and 20 Gyr 
(solid); exponentially increasing SFR with
e-folding time of 1 Gyr, started 10 Gyr ago (dashed).}
\end{figure}

A variety of DTDs have bee proposed in the literature, some empirical, 
others derived from stellar evolution theory. Fig. 1 shows the
selection of functions considered here.
The MDP (\cite{MDP}) distribution accomodates two
channels, one \textit{prompt} and one \textit{delayed}: the first includes
events with  delay times shorter than 0.1 Gyr; the second populates the 
remaining, very wide range of delay times up to one Hubble time, with
a flat distribution between 0.1 and 3 Gyr, plus a progressive (rapid) depletion
of events at longer delays. The time span and strength of the \textit{prompt} 
channel are constrained by the controversial evidence of an enhanced SNIa 
rate in radio loud ellipticals.
The S04 (\cite{Strolger}) distribution allows for a narrow
range of delay times: this DTD appeared to provide the best fit of the
evolution of the cosmic SNIa rate with redshift in combination with the
cosmic SFR by \cite{Giava}.  
The interpretation of the data which leads to these DTDs has been recently 
questioned by \cite{GRD}.
The GR05 (\cite{GR}) models are rooted on stellar 
evolution, and are meant to describe a general shape for the DTD which
follows from the characteristics of the clock of the explosion, and
the distribution of the binary parameters in the progenitors' systems. 
In the Single Degenerate (SD) model a SNIa occurs when a White Dwarf (WD)
accreting from an evolved companion reaches the Chandrasekhar mass;
the Double Degenerate (DD) model corresponds to close binary WDs 
progenitors with
total mass exceeding the Chandrasekhar limit, which merge due to  
gravitational waves radiation losses. The WIDE and CLOSE variety differ for the
efficiency with which the first common envelope phase shrinks the orbit of the
progenitors: this efficiency is higher in the CLOSE than in the WIDE case,
leading to binary WDs with smaller separations, and then to
a DTD characterized by a higher fraction of short delay times.
The minimum mass of the
secondary in the progenitor system, and the distribution of the separations of
the DD systems at birth concur to shape the DTD for the DDs: here we choose two
extreme cases, i.e. a steep DD CLOSE, and a flat DD WIDE model.
 
All the GR05 models encompass a wide range of delay times, reflecting 
either the evolutionary lifetimes of secondaries from 8 down to 0.8 M$_\odot$ 
(SD models), or the wide range of gravitational wave radiation delays 
ensuing from a range of separations of the DD systems. 
At the same time, all the models are more populated at short, 
rather than long delays, basically due to the higher
evolutionary rate off the main sequence of massive stars, compared
to low mass stars. The late epoch drop of the DTD for the SD model is caused
by the requirement of making up the Chandrasekhar mass in systems with
progressively less massive secondaries. The smooth shapes of the GR05 DTDs
reflect the assumption of continuous distributions  
of the binary parameters in SNIa progenitors, between wide limits. 
Conversely, the S04 DTD requires a very narrow range of these 
parameters. Compared to the MDP function, the GR05 models lack very early 
events. The minimum
delay time for the latter models is fixed by the evolutionary lifetime of 
the most massive secondary in the progenitor systems, which is
not well constrained. A maximum mass of 8 (10) M$_\odot$
implies a minimum delay time of $\simeq 36 (25)$ Myr, if core overshooting
occurs during evolution; or $\simeq 32$ Myr for a standard (no overshoot)
8 M$_\odot$ track. The GR05 models adopt a minimum delay time of 40 Myr, 
but a shorter delay could be applicable. The main difference
between the MDP and the GR05 DTDs is the 
sharp drop in the former at a delay time of 0.1 Gyr, which corresponds to an 
evolutionary mass of $\simeq$ 5 M$_\odot$. In the GR05 models nothing special 
happens when the secondary mass drops below 5 M$_\odot$, as   
the distribution of the binary parameters (in SNIa progenitor systems) 
is assumed to be  continuous across this mass. 

The MDP and GR05 DTDs imply that the younger the stellar 
system, the higher its SNIa rate per unit mass, since these DTDs favour
early, rather than late, events. The S04 DTD, instead, predicts a maximum
rate in systems with a high fraction of intermediate age stars.
To proceed with a quantitative fit of the specific SNIa rate as a function
of the parent galaxy properties we need to specify the age distribution of
stars in galaxies.
Following \cite{Gavazzi} we consider the family of SFRs illustrated as
solid lines in the right panel of Fig. 1. 
We assumed that all galaxies started to
form stars 13 Gyr ago and are currently active, but, going from early to
late type galaxies, the peak of the SF
activity moves to younger ages, and the age distribution becomes wider.
The results of the 
convolution of the DTDs with this family of SFH is presented in the next 
section. A SFH skewed towards the current epoch is also 
considered, to reproduce the colors of the bluest galaxies (in $B-K$).

\section{Fitting the observed SNIa rates in nearby galaxies}

\begin{figure}
\includegraphics[height=.3\textheight]{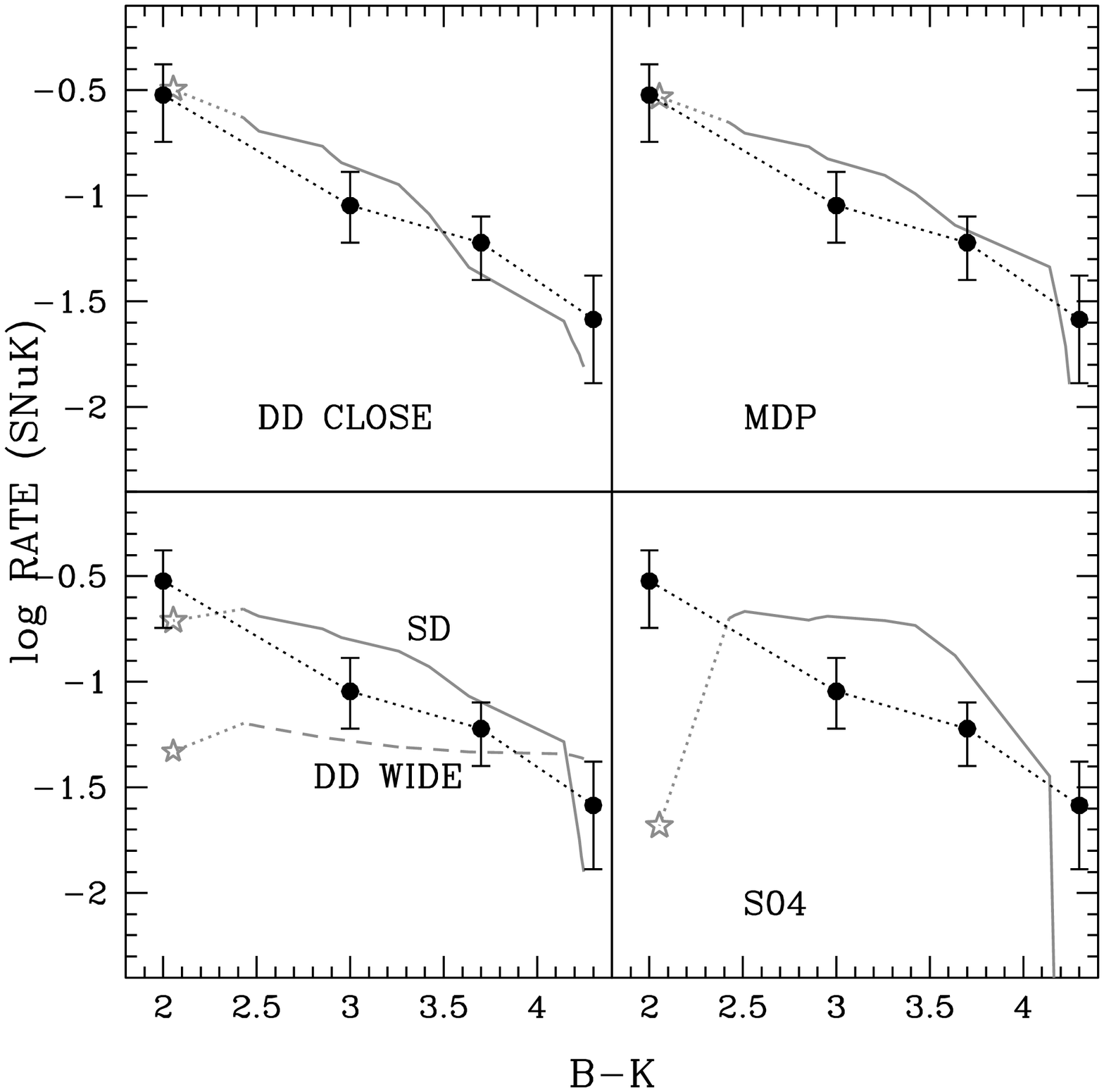}
\includegraphics[height=.3\textheight]{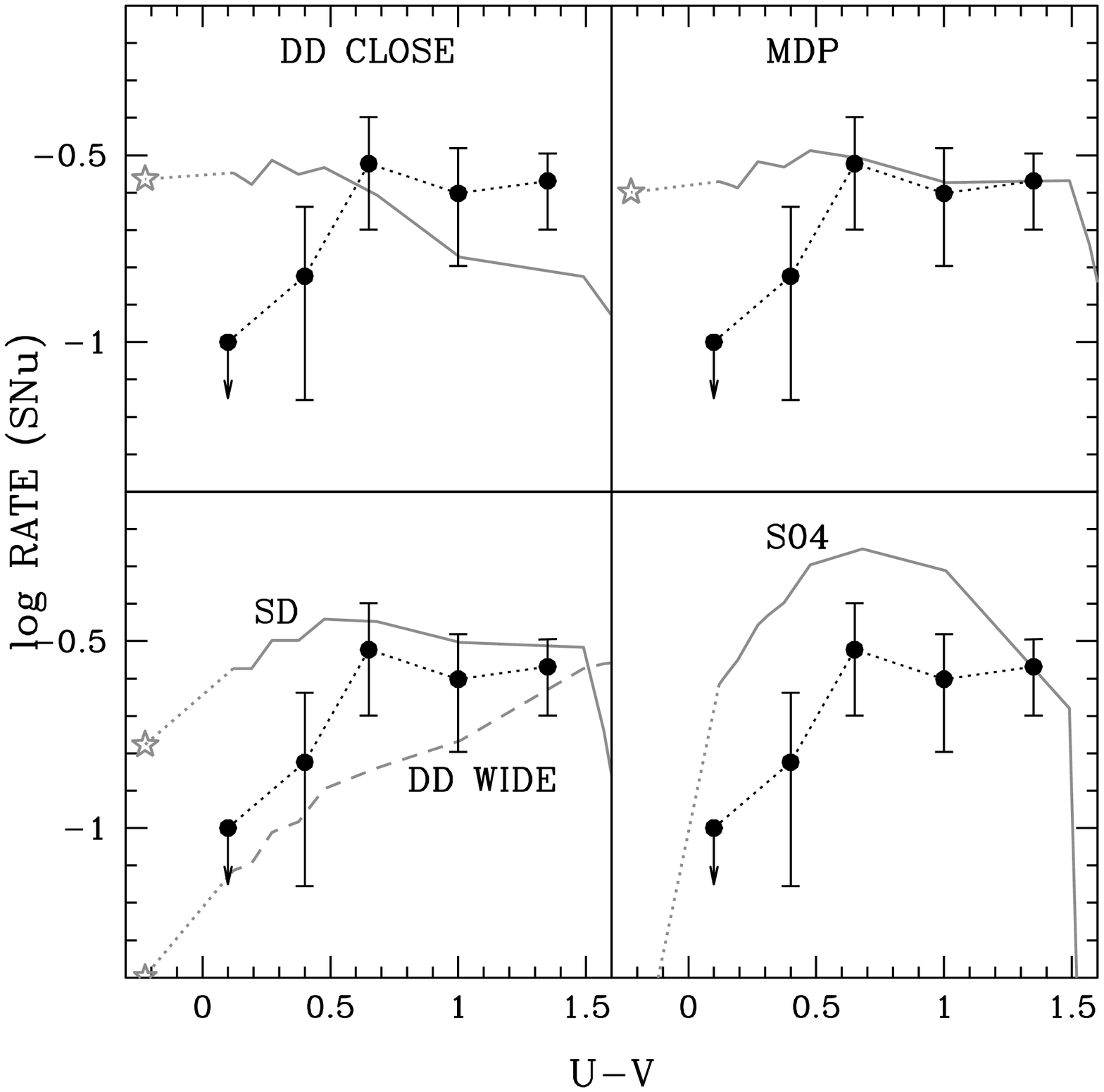}
  \caption{SNIa rates as function of the parent galaxy colors per unit 
$K$-band (left) and $B$-band (right) luminosity. Observational determinations 
are plotted as filled dots together with their error bars.
The grey lines are models obtained with the DTDs (as labelled) and
SFRs in Fig. 1. The star shows the prediction of the exponentially increasing
SFR. The models assume a SNIa
productivity $k_{\rm Ia}$= 0.001 and 0.003 M$_\odot^{-1}$ for the DD WIDE 
and all others models, respectively. The metallicity 
is assumed to correlate with $\tau_{\rm SF}$; specifically: $Z$=0.02 for 
$\tau_{\rm SF} \leq 3$Gyr; $Z$=0.008 for 3 Gyr $< \tau_{\rm SF} \leq$ 6 Gyr; 
$Z$=0.004 for
6 Gyr $< \tau_{\rm SF} \leq$ 9 Gyr; $Z$=0.001 for $\tau_{\rm SF}> 9$ Gyr and 
for the exponentially increasing SFR.} 
\end{figure}

The full dots in Fig. 2 show the SNIa rate per unit luminosity as a function
of the color of the parent galaxy from \cite{Mannu05} (left panel)
and from \cite{CET99} (right panel). Rather than
considering the SNIa rate per unit stellar mass, as done in 
\cite{MDP}, we prefer to fit the rate per unit luminosity which is
a fully observational quantity, whereas the rate per unit mass rests upon 
a relation between the mass-to-light ratio and the galaxy color 
based on stellar population models. It can be noticed that, while the rate per
unit $K$ band luminosity is much larger in blue (young) galaxies than in red
(old) ones, the rate per unit $B$ band luminosity is lower in the bluest
objects, and it is virtually constant in galaxies with $U-V \geq 0.5$.
It is generally believed that $L_{\rm K}$ traces mass better than $L_{\rm B}$:
the data in the left panel of Fig. 2 would imply that, going from old to 
young stellar systems, the rate per unit mass increases, as predicted by 
the MDP and GR05 DTDs.
At the same time, the $L_{\rm B}/M$ ratio is larger in young systems, 
so that the trend of the rate in SNu's depends on whether $L_{\rm B}/M$ 
increases
faster than the rate per unit mass as the average age of the stellar population
becomes younger. The data in the right panel in Fig. 2 seem to indicate 
that at $U-V > 0.5$ the two factors scale in the same way, while
in the bluest galaxies the increase of the $L_{\rm B}/M$ ratio is the 
dominating
effect. The grey lines in Fig. 2 show the theoretical relations as given by 
Eq. (1) for the DTDs and SFHs plotted on Fig. 1. The colors of the model 
galaxies result from population synthesis computations based on 
simple stellar populations (SSP) models by \cite{Marigo}.
A trend of decreasing metallicity 
with increasing $\tau_{\rm SF}$ has been adopted to 
construct the galaxy colors. In spite of the low metallicity, the $B-K$
color of the model with $\tau_{\rm SF}$=20 Gyr is redder than the bluest 
bin of data in \cite{Mannu05}. 
This color can be reproduced by assuming a SFR strongly peaked at the current
epoch: the star indicates the model predictions obtained with the 
exponentially increasing SFR shown in Fig. 1.
Similar to previous conclusions (e.g. \cite{MDP}, \cite{GR}), the comparison 
between models and observations of the
rate in SNuK indicates that the DTD needs to be mostly populated at the short 
delay times: the DD WIDE model predicts a flat specific rate with the 
parent galaxy 
color, and the S04 DTD implies a maximum rate in galaxies with intermediate
average age, hence intermediate colors. None of the models, however, seems 
able to fit the
trend of the rate in SNu's with the parent galaxy $U-V$ color. 
The increase of the rate per unit mass from intermediate age to young
galaxies predicted by the DD CLOSE and the MDP curves matches the increase
of the $L_{\rm B}/M$ ratio, so that the latest galaxy types should have about
the same SNIa rate in SNu's as spiral galaxies. For the other DTDs we
do obtain a lower rate in the bluest galaxies, but the overall trend of the
data is not well reproduced. Notice that on Fig. 2 we compare only
the shape of the relation between the SNIa rate and the parent galaxy color:   
a vertical shift of the models is allowed, given that 
the productivity $k_{\rm Ia}$ is a free parameter.
Nevertheless, the same DTD does not fit the trend of the SNIa rate with
the parent galaxy color on both observational planes.
This conflict could be originated by several causes, including inadequacies
in the adopted SFH laws and/or in the population synthesis models. In
addition, although the global galaxy sample is the same, the observational 
relations in the two panels are not constructed for the same individual 
galaxies, as will be illustrated later. Computation
of models with a broad family of SFHs, which include exponential and power
law distributions, and with formation epochs varying between 3 and 13 Gyr ago
lead to results very similar to those shown in Fig. 2. This does not exclude
that some ad hoc variation of the SFH across the galaxy sample may lead 
to a unique indication on the DTD, but we clearly need to better constrain
the age distribution of the stars in the various galaxy types which are 
used to build the observational relation. This is attempted next.
   
\begin{figure}
\includegraphics[height=.35\textheight]{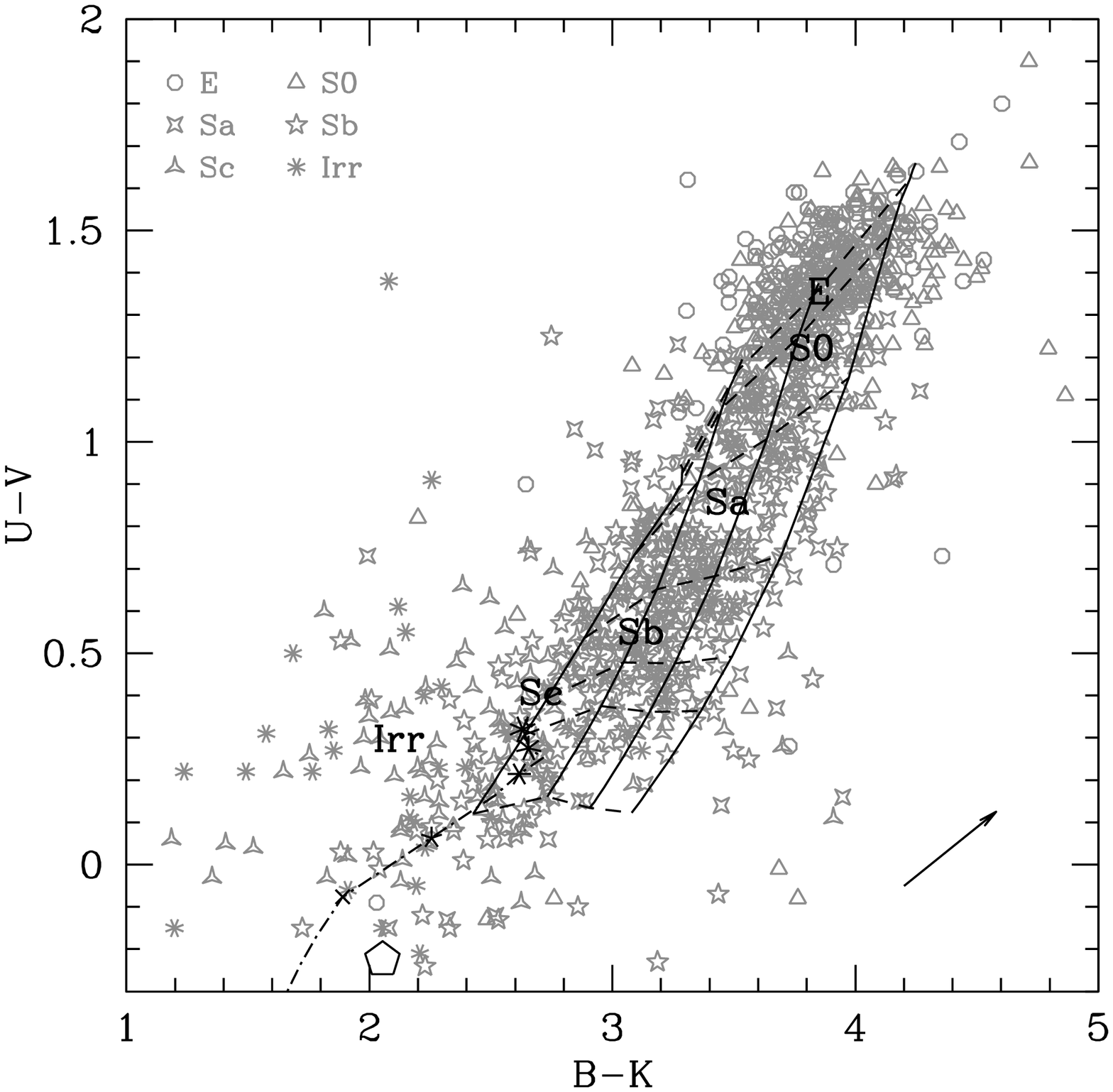}
\includegraphics[height=.35\textheight]{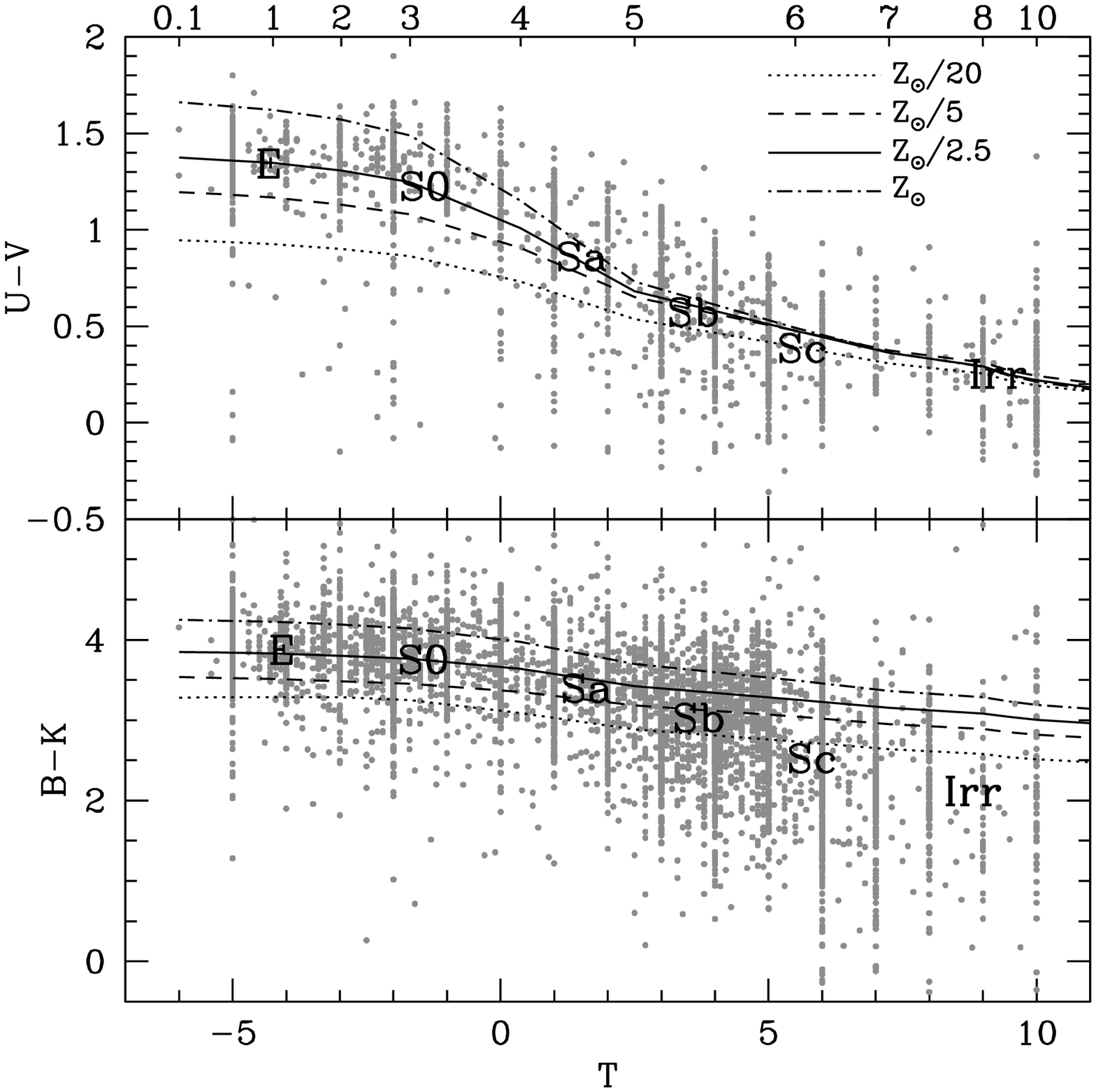}
\caption{Calibration of the parameter $\tau_{\rm SF}$ versus galaxy type $T$.
Left: two color diagram for the galaxies (no reddening correction applied); 
the symbols encode the galaxy type, as labelled, and the letters are located
at the average colors of the respective galaxy type. The solid 
lines show models with
$\tau_{\rm SF}$ increasing from 0.1 to 20 Gyr and a fixed metallicity of
$Z=0.001,0.004,0.008,0.02$ (from left to right). The dashed lines 
connect models of different metallicity and fixed $\tau_{\rm SF}$
=(1,3,4,5,6,7,20) Gyr (from top to bottom). The big pentagon shows the 
colors of a model
with an exponentially increasing SFR (e-folding time of 1 Gyr), 
and metallicity Z=0.001. The arrow shows the reddening vector. 
The dot dashed line illustrates the effect of a recent burst
of star formation, involving 10\% of the total mass, superimposed on a 
young galaxy model (both components with $Z$=0.001); the dots show the colors
for a burst age of (0.1,0.15,0.3,0.5,1) Gyr. Right:
the colors are plotted as function of $T$ for individual galaxies; the letters
are located at the average color and $T$ of the respective galaxy type.
Lines show models as function of $\tau_{\rm SF}$ (labelled on the top x-axis) 
for various metallicities, as labelled.}
\end{figure}

The galaxy sample from \cite{CET99} includes 9346 galaxies from the RC3 
catalogue, which provides $B$ magnitudes for all of them, morphological type 
$T$ for most
(actually 9279 galaxies), and $U-B$, $B-V$ only for a subset.
The $U-V$ color can be computed only for 1796
galaxies of the sample, which may be a small number to derive a robust
relation between the SNIa rate with the properties of the parent galaxy.
Cross correlating this galaxy sample with 2MASS yields infrared magnitudes 
for 6562 objects, for all of which the $B-K$ color can be computed. Although
this is a large sample, the maximum exploitation of the data base is 
realized by examining the SNIa rate as a function of the parent galaxy 
morphological type. To do that we need to relate $T$
to the SFH in the different galaxies.
The model colors computed for the family of SFHs in Fig. 1 and 
different values of the metallicity (fixed metallicity for each SFH) are 
shown as black lines in the left diagram of Fig. 3, superimposed on the colors 
of the 1108 galaxies of the sample for which both $U-V$ 
and $B-K$ are available.
These models encompass most of the data, and fairly describe the average 
trend of both colors
becoming progressively redder with the star formation timescale becoming 
shorter. It also seems that the later galaxy types require lower 
metallicity compared to the earlier types, which is reasonable, given the
low metallicity in the interstellar medium of the latest galaxy types. 
However, there's
a substantial number of galaxies which appear too blue in $B-K$ for
their $U-V$ with respect to the models, especially among 
the latest types, for which even the average colors
fall outside the models grid.
We have checked a few possibilities to 
account for the location of the Irregular galaxies on this plot, including the
effect of reddening, of different SSP models (e.g. \cite{CLAUDIA}), 
of recent bursts of SF. None of these options leads
to a fair fit of the colors of these galaxies. In particular, the colors of
a recent burst are blue in both $B-K$ and $U-V$, as shown by the model with 
the exponential SFR, so that the addition of a recent burst does not 
populate the region occupied by the Irregulars.
Notwithstanding this problem, we proceed with the calibration
of the $\tau_{SF}$ parameter versus $T$ by comparing the average colors of
the various galaxy types with the models. This yields the correspondence 
shown in the right diagram on Fig. 3, which is very similar to what found by
\cite{Gavazzi} from the comparison of the whole (average) 
spectral energy distributions of galaxies in the Virgo cluster with 
models based on this
same SFHs and SSP models by \cite{BC93}. The right diagram on 
Fig. 3 shows that this calibration is suitable for all galaxies with $U-V$
color (1796), and acceptable for all galaxies with $B-K$ color (6562), in the 
sample of \cite{CET99}. The data 
show a sizeable scatter with respect to the models, especially for late type 
galaxies, but the average colors do lie on the theoretical lines, 
with the exception of the Irregulars, which 
again appear too blue in $B-K$ even for the lowest metallicity models. 

Fig. 4 shows the SNIa rates as function of the morphological type 
compared to the predictions from Eq. (1) with the 5 selected DTDs,
having adopted the $\tau_{\rm SF}$ vs $T$ calibration described above.
The observed rates have been re-determined with respect to previous work, to
incorporate the same partition of galaxy types as used in the calibration.
In particular, the early type class (E+S0), whose rate is shown as a star,
has been split into two separate classes. Within the large error bars, 
the two diagrams now yield the same conclusion with respect to the DTDs:
the S04 function is ruled out because it highly under-predicts events in Es 
and S0s; the DD WIDE function is also ruled out because it over-predicts events
in such galaxies; all other DTDs are instead compatible with the data.
We notice that if the drop of the SNIa rate per unit $B$ band luminosity going
form S0 to Es is real, the MDP and SD models are favoured, since the DD
models are featureless at long delay times. This drop is not appreciable
when E and S0 galaxies are combined in one class.

\begin{figure}
\includegraphics[height=.35\textheight]{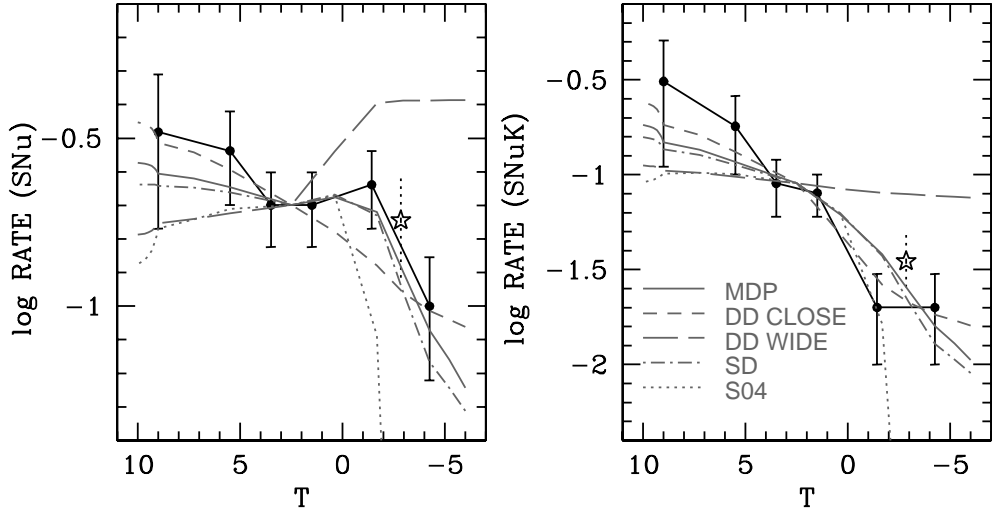}
  \caption{SNIa rates as function of the galaxy morphological type. Black 
points show the observational determinations; the star is the rate determined
by combining the data for the E and S0 in one single class. Grey lines show 
the models shifted to reproduce the observed rate in Sb galaxies}
\end{figure}

The models in Fig. 4 have been shifted to reproduce the observed rate in Sb
galaxies; the shift yields a value for the productivity $k_{\rm Ia}$. For a 
Salpeter initial mass function, flattened below 0.5 M$_\odot$, the productivity
turns out $k_{\rm Ia} \simeq$ (3,1.5,2, 2.3,1.3)\, $10^{-3}$ M$_\odot^{-1}$ 
respectively
for the DD CLOSE, DD WIDE, SD, MDP and S04 models, with little variation 
when considering the rates in SNu's or in SNuK's. This same IMF has
$\sim 0.06$ M$_\odot^{-1}$ stars with mass between 2 and 8 M$_\odot$. 
Therefore, globally, its suffice that $\sim 4 \%$ of these stars end 
their lives as SNIa. The MDP DTD requires that $\sim$ 50\% of the
events originate from short lived progenitors, with masses larger than
$\sim$ 5 M$_\odot$. One stellar generation provides $\sim 0.01$ M$_\odot^{-1}$ 
stars with mass between 5 and 8 M$_\odot$. If $50 \%$ of the events in the 
calibrating Sb galaxies
are produced by the \textit{prompt} channel, its efficiency of SNIa production
should be $\sim 0.5 \cdot 0.23 \simeq 0.12$, i.e. larger than
in the other cases, but still acceptable. 

\section{Conclusions}

The indications on the DTD from the fit of the specific SNIa rate as 
function of the parent galaxy colour appear inconsistent when considering 
different colors to trace the SFH and the $M/L$ ratio. To overcome this 
problem, the age
distribution of galaxies in the sample of \cite{CET99} has been constrained
on the two color plot which accounts for both $U-V$ and $B-K$ colors. 
In this way, it is possible to derive a calibration of the morphological
type in terms of SFH, and proceed with fitting the trend of the specific
SNIa rate as function of the morphological type. On this plane, the specific
rates measured in SNu's and in SNuK's provide the same indication for the 
DTD, as
well as similar values for the SNIa productivity from one stellar generation
($\sim 0.002-0.003 M_{\odot}^{-1}$) .
The constraints on the DTD are rather broad: the S04 distribution and a 
very flat DD WIDE model are ruled out, but all other models are acceptable.
We notice that the parameters for the DD models have been intentionally
chosen so as to maximize the difference between DD CLOSE and DD WIDE cases,
and we anticipate that  
flatter DD CLOSE and steeper DD WIDE models are also acceptable.
Our modelling does not reproduce the average $B-K$ of Irregular
galaxies; however 
our conclusions on the DTD
do not rely on the rate measured in these galaxies, but rather 
on the trend of the SNIa rate in the other morphological types. 
Stronger constraints on the age distributions in the various galaxies in
the sample 
(e.g. from the analysis of their full spectral energy distributions),
especially for the latest galaxy types, will enable us to derive
stronger constraints on the DTD. In addition, a robust measurement of the 
rate in E and S0s separately will be very important to verify whether the
DTD features a sizable drop at the long delays.

\end{document}